\documentclass [12pt]{article}
\usepackage{amsmath,amssymb,cite}

\usepackage{float}
\usepackage[section]{placeins}
\usepackage{tabularx}
\usepackage[english]{babel}
\usepackage[dvips]{graphicx}
\usepackage{indentfirst}
\usepackage{epsfig}
\numberwithin{equation}{section}

\begin{document}

\title{
Emergent geometry from random multitrace matrix models}

\author{B. Ydri\footnote{Email:ydri@stp.dias.ie}, A. Rouag, K. Ramda\\
Department of Physics, Faculty of Sciences, Annaba University,\\
 Annaba, Algeria.
}
\maketitle
\abstract{A novel scenario for the emergence of geometry in random multitrace matrix models of a single hermitian matrix $M$ with unitary $U(N)
$ invariance, i.e. without a kinetic term, is presented. In particular, the dimension of the emergent geometry is determined from the critical exponents of the disorder-to-uniform-ordered transition whereas the metric is determined from the Wigner semicircle law behavior of the eigenvalues distribution of the matrix $M$. If the uniform ordered phase is not sustained in the phase diagram then there is no emergent geometry in the multitrace matrix model. 
}


\section{Introduction and Motivation}

The original motivation for this work is the theory of noncommutative $\Phi^4$ which we now briefly describe. A scalar phi-four theory on a non-degenerate noncommutative Euclidean spacetime is a  a three-parameter matrix model of the generic form
\begin{eqnarray}
S&=&{\rm Tr}_H\big(aM\Delta M+bM^2+cM^4\big).\label{fundamental}
\end{eqnarray} 
The Laplacian $\Delta$ captures precisely the underlying geometry, i.e. the metric, of the  noncommutative Euclidean spacetime in the sense of  \cite{Connes:1994yd,Frohlich:1993es}. This theory can be regularized non-perturbatively using $N\times N$ matrices in an almost obvious way, i.e. the Hilbert space $H$ can be taken to be finite dimensional of size $N$. This theory exhibits the following three known phases: 
\begin{itemize}
\item The usual $2$nd order Ising phase transition between disordered $<M>=0$ and uniform ordered $<M>\sim {\bf 1}_N$ phases. This appears for small values of $c$. This is the only transition observed in commutative phi-four, and thus it can be accessed in a small noncommutativity parameter expansion.
\item  A matrix transition between disordered $<M>=0$ and non-uniform ordered $<M>\sim \gamma$ phases with $\gamma^2={\bf 1}_N$. This transition coincides, for very large values of $c$,  with the $3$rd order transition of the real quartic matrix model, i.e. the model with $a=0$, which occurs at $b=-2\sqrt{Nc}$. In terms of $\tilde{b}=bN^{-3/2}$ and $\tilde{c}=cN^{-2}$ this reads
\begin{eqnarray}
\tilde{b}=-2\sqrt{\tilde{c}}.\label{cl}
\end{eqnarray} 
This is therefore a transition from a one-cut (disc) phase to a two-cut (annulus) phase \cite{Brezin:1977sv,Shimamune:1981qf}. See also \cite{eynard, Kawahara:2007eu}.
\item  A transition between uniform ordered  $<M>\sim {\bf 1}_N$ and non-uniform ordered $<M>\sim \gamma$ phases.  The non-uniform phase, in which translational/rotational invariance is spontaneously broken, is absent in the commutative theory. The non-uniform phase is essentially  the stripe phase observed originally on  Moyal-Weyl spaces in \cite{Gubser:2000cd,Ambjorn:2002nj}. 
\end{itemize}
Thus, the  uniform ordered phase $<\Phi>\sim {\bf 1}_N$ is stable in the theory (\ref{fundamental}). This fact is in contrast with the case of the
real quartic matrix model  $V={\rm Tr}_H({b}M^2+{c}M^4)$ in which this solution becomes unstable for all values of the couplings. The source of this stability is obviously the addition of the kinetic term to the action.

 The non-uniform ordered phase \cite{brazovkii} is a full blown nonperturbative manifestation of the perturbative  UV-IR mixing effect \cite{Minwalla:1999px} which is due to the underlying highly non-local matrix degrees of freedom of the noncommutative scalar field.

The above picture of the phase diagram holds for noncommutative phi-four in any dimension, and the three phases are all stable, and are expected to meet at a triple point. The phase structure in four dimensions was discussed using the Hartree-Fock approximation  in \cite{Gubser:2000cd} and studied by means of the Monte Carlo method, employing the fuzzy torus \cite{Ambjorn:2000cs} as regulator, in \cite{Ambjorn:2002nj}.

In two dimensions the noncommutative phi-four theory is renormalizable \cite{Grosse:2003nw}. The regularized theory on the fuzzy sphere \cite{Hoppe:1982,Madore:1991bw} is given by the action (\ref{fundamental}) with a finite dimensional Hilbert space $H$ of size $N$ and a Laplacian $\Delta=[L_a,[L_a,..]]$ where $L_a$ are the generators of $SU(2)$ in the IRR of spin $(N-1)/2$. 

\begin{figure}[htbp]
\begin{center}
\includegraphics[width=5.0cm,angle=-90]{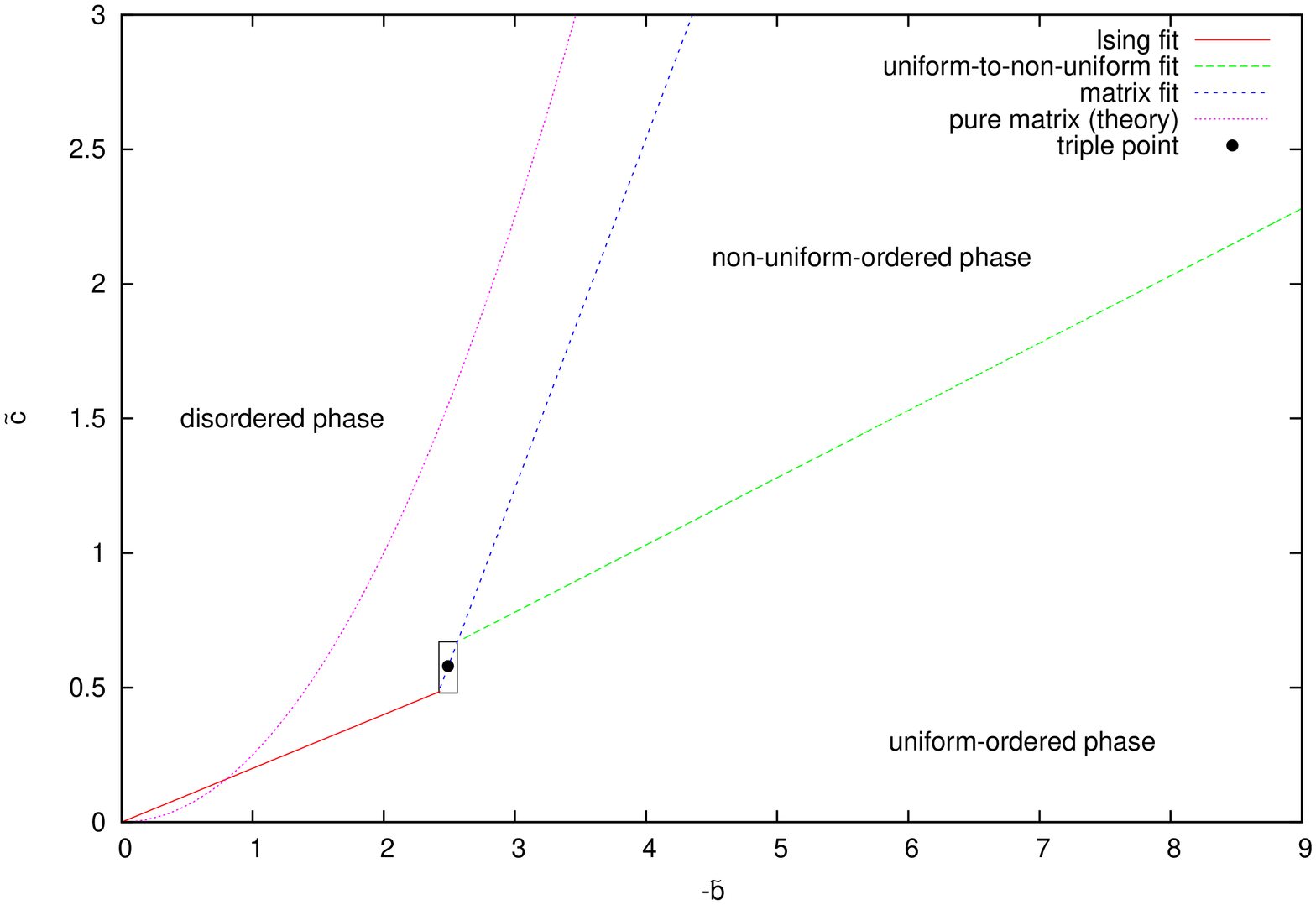}
\includegraphics[width=5.0cm,angle=-90]{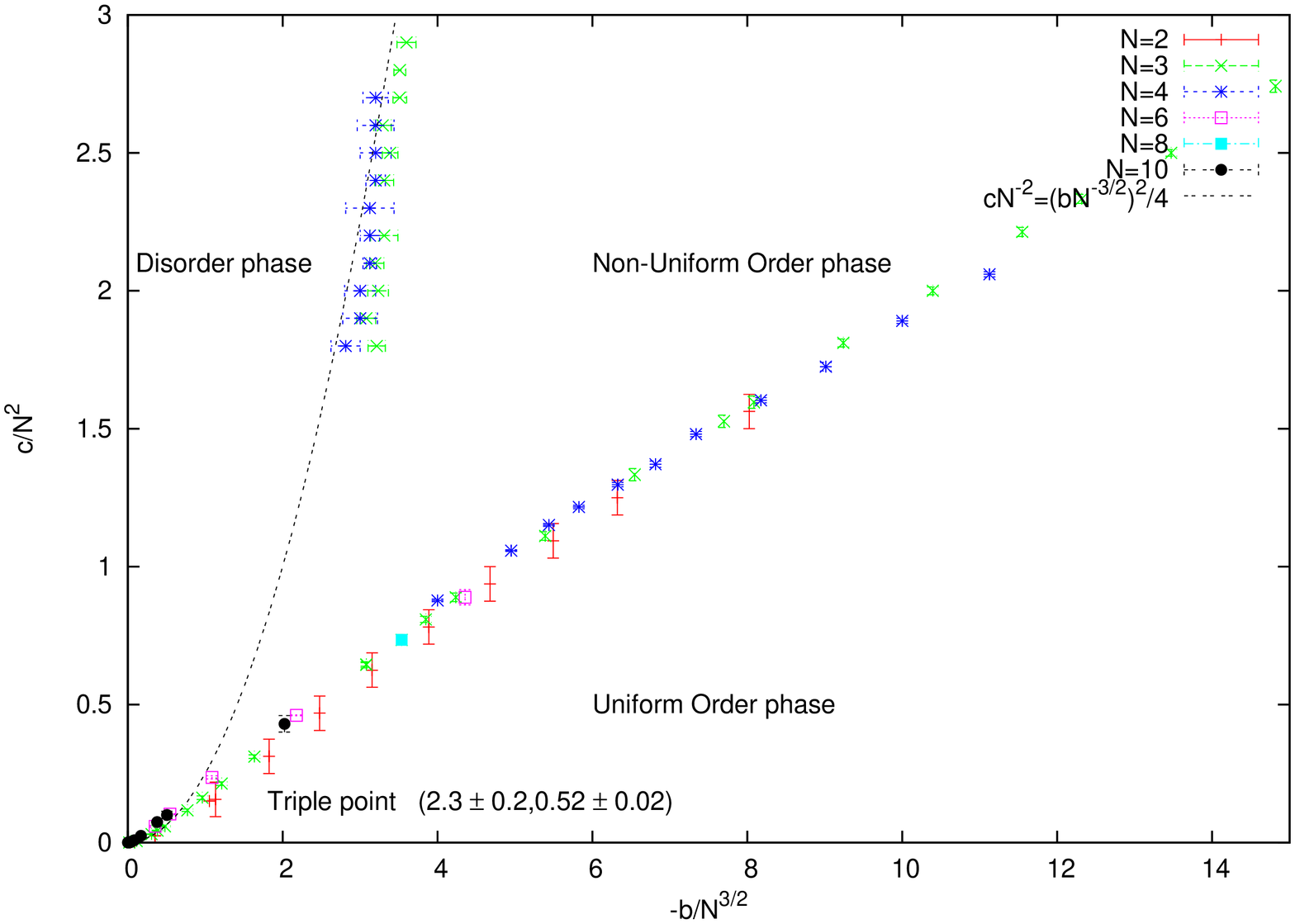}
\caption{The phase diagram of noncommutative phi-four theory on the fuzzy sphere. In the first figure the fits are reproduced from actual Monte Carlo data \cite{Ydri:2014rea}. Second figure reproduced from \cite{GarciaFlores:2009hf} with the gracious permission of  D.~O'Connor.}\label{phase_diagram}
\end{center}
\end{figure}

The above phase structure was confirmed in two dimensions by means of Monte Carlo simulations on the fuzzy sphere in  \cite{GarciaFlores:2009hf,GarciaFlores:2005xc}. Indeed, fuzzy scalar phi-four theory enjoys three stable phases: i) disordered (symmetric, one-cut, disk) phase, ii) uniform ordered (Ising, broken, asymmetric one-cut) phase and iii) non-uniform ordered (matrix, stripe, two-cut, annulus) phase. The phase diagram is shown on the two graphs of figure (\ref{phase_diagram}) which were generated using the Metropolis algorithm. 

The problem of the phase structure of fuzzy phi-four was also studied by means of the Monte Carlo method in \cite{Martin:2004un,Panero:2006bx,Medina:2007nv,Das:2007gm,Ydri:2014rea}. The analytic derivation of the phase diagram of noncommutative phi-four on the fuzzy sphere was attempted in \cite{O'Connor:2007ea,Saemann:2010bw,Polychronakos:2013nca,Tekel:2014bta,Nair:2011ux,Tekel:2013vz,Ydri:2014uaa,Steinacker:2005wj}. 

The related problem of Monte Carlo simulation of noncommutative phi-four on the fuzzy torus, and the fuzzy disc was considered in \cite{Ambjorn:2002nj}, \cite{Bietenholz:2004xs}, and \cite{Lizzi:2012xy} respectively. For a recent study see \cite{Mejia-Diaz:2014lza}.

In \cite{Ydri:2014rea} the phase diagram of fuzzy phi-four theory was computed by Monte Carlo sampling of the eigenvalues $\lambda_i$ of the scalar field $M$. This was possible by coupling the scalar field $M$ to a $U(1)$ gauge field $X_a$ on the fuzzy sphere which then allowed us, by employing the $U(N)$ gauge symmetry, to reduce scalar phi-four theory to only its eigenvalues. 
 The pure gauge term is such that the gauge field $X_a$ is  fluctuating around $X_a=L_a$. 

Another powerful method which allows us to reduce noncommutative scalar phi-four theory to only its eigenvalues, without the additional dynamical gauge field, is the multitrace approach.
The multitrace approach was initiated in \cite{O'Connor:2007ea,Saemann:2010bw}. See also \cite{Ydri:2014uaa} for a review and an extension of this method to the noncommutative Moyal-Weyl plane. For an earlier approach see \cite{Steinacker:2005wj} and for a similar more non-perturbative approach see \cite{Polychronakos:2013nca,Tekel:2014bta,Nair:2011ux,Tekel:2013vz}. The  multitrace expansion is the analogue of the Hopping parameter expansion on the lattice in the sense that we perform a small kinetic term expansion, i.e. expanding in the parameter $a$ of (\ref{fundamental}), while treating the potential exactly. This should be contrasted with the small interaction expansion of the usual perturbation theory. The effective action obtained in this approach is a matrix model which can be expressed solely in terms of the eigenvalues $\lambda_i$ and which, on general grounds, can only be a function of the combinations $T_{2n}\propto \sum_{i\neq j}(\lambda_i-\lambda_j)^{2n}$. To the lowest non-trivial order we get an effective action of the form  \cite{Ydri:2014uaa,Polychronakos:2013nca, Saemann:2010bw}


\begin{eqnarray}
S_{\rm eff}&=&\sum_{i}(b\lambda_i^2+c\lambda_i^4)-\frac{1}{2}\sum_{i\neq j}\ln(\lambda_i-\lambda_j)^2\nonumber\\
&+&\bigg[\frac{aN}{4}v_{2,1}\sum_{i\ne j}(\lambda_i-\lambda_j)^2+\frac{a^2N^2}{12}v_{4,1}\sum_{i\ne j}(\lambda_i-\lambda_j)^4-\frac{a^2}{6}v_{2,2}\big[\sum_{i\ne j}(\lambda_i-\lambda_j)^2\big]^2+...\bigg].\label{ft}\nonumber\\
\end{eqnarray}
The logarithmic potential arises from the Vandermonde determinant, i.e. from diagonalization. The coefficients $v_{2,1}$, $v_{4,1}$ and $v_{2,2}$ are given by  $v_{2,1}=+1~,~v_{4,1}=0~,~v_{2,2}={1}/{8}$. Furthermore, it is not difficult to convince ourselves that the above action is a multitrace matrix model since it can be expressed in terms of various moments $m_n=Tr M^n$ of the matrix $M$. 

The original multitrace matrix model written down \cite{O'Connor:2007ea} comes with different values of $v$'s and therefore, in the commutative limit $N\longrightarrow \infty$, it corresponds to a phi-four theory on the sphere modulo multi-integral terms. 

Since these multitrace matrix models depend only on $N$ independent eigenvalues their Monte Carlo sampling by means of the Metropolis algorithm does not suffer from  any ergodic problem. The phase diagrams of these models obtained in Monte Carlo simulations will be reported elsewhere.  

The remainder of this article is organized as follows:
\begin{enumerate}
\item{}Section $2$: We describe our proposal for how fuzzy geometry can emerge in generic multitrace matrix models.
\item{}Section $3$: We apply our proposal to an explicit example. We will show that if the multitrace matrix model under consideration does not sustain the uniform ordered phase then there is no emergent geometry. On the other hand, if the uniform ordered phase is sustained then there is an underlying or emergent geometry. In particular, we will show how 
\begin{itemize}
\item i) to determine the dimension from the critical exponents of the uniform-to-disordered (Ising) phase transition, and how
\item{}
ii) to determine the metric (Laplacian, propagator) from the Wigner semicircle law behavior of the eigenvalues distribution of the matrix $M$. 
\end{itemize}
\item{}Section $4$: We conclude by giving a straightforward generalization to fuzzy ${\bf CP}^n$ and fuzzy ${\bf T}^n$.
\end{enumerate}

\section{The Proposal}
We start with a general multitrace matrix model rewritten in terms of the moments  $Tr M^n$  with generic parameters $B$, $C$, $D$,  $B^{'}$, $C^{'}$, $D^{'}$,  $A^{'}$,...as
 \begin{eqnarray}
V&=&{B}Tr M^2+{C}Tr M^4+D\bigg[ Tr M^2\bigg]^2\nonumber\\
&+&B^{'} (Tr M)^2+C^{'} Tr M Tr M^3+D^{'}(Tr M)^4+A^{'}Tr M^2 (Tr M)^2+....\label{fundamental1}
\end{eqnarray}
This action includes the noncommutative phi-four model on the fuzzy sphere (\ref{ft}) and the multitrace matrix model of \cite{O'Connor:2007ea} as special cases. It also includes as special cases the multitrace matrix models obtained by expanding the kinetic term  on i) fuzzy ${\bf CP}^n$ \cite{Balachandran:2001dd,Saemann:2010bw}, on ii) Moyal-Weyl spaces with and without the harmonic oscillator term \cite{Ydri:2014uaa}, and on iii) fuzzy tori \cite{Ambjorn:2000cs}. 

The phase diagram of the action (\ref{fundamental1}) will generically contain the matrix one-cut-to-two-cut transition line separating the two stables phases of disorder and non-uniform-order. However, the uniform ordered phase will typically be unstable as in the case of the real quartic matrix model 
 \begin{eqnarray}
V&=&{B}Tr M^2+{C}Tr M^4.\label{fundamental2}
\end{eqnarray}
Our proposal goes as follows. We can check for a possible emergence of geometry in the multitrace matrix model (\ref{fundamental1}) by following the three steps:
\begin{enumerate}
\item We compute the phase diagram of the model (\ref{fundamental1}). If the uniform ordered phase remains unstable as in the case of the real quartic matrix model (\ref{fundamental2}) then there is no geometry and the model is just a trivial deformation of  (\ref{fundamental2}). In the opposite case we claim that there is an underlying, i.e. emergent, geometry with a well defined dimension (step $2$) and a well defined Laplacian/metric (step $3$). This means that we can rewrite the multitrace matrix model, in the region of the phase diagram where the uniform ordered phase exists,  in terms of a scalar function and a star product with a noncommutativity parameter $\theta$ by finding the appropriate Weyl map. As a consequence, a small noncommutativity parameter expansion can be performed and the the limit $\theta\longrightarrow 0$ can be taken. The disordered-to-uniform-ordered phase transition reduces therefore to the usual $2$nd order Ising phase transition on the underlying geometry. 
\item  We compute the dimension of the underlying by computing the critical exponents of the  disordered-to-uniform-ordered phase transition which, by universality, take specific values in each dimension.
\item We compute the Laplacian by computing the free behavior of the propagator. This is done explicitly by computing the eigenvalues distribution of the matrix $M$ in the free regime, small values of $C$, and comparing with the Wigner semicircle law behavior which must hold with a specific radius depending crucially on the kinetic term.
\end{enumerate}

\section{Explicit Example: The Fuzzy Sphere}
\subsection{Phase Diagram}
\begin{figure}[htbp]
\begin{center}
\includegraphics[width=9.0cm,angle=-90]{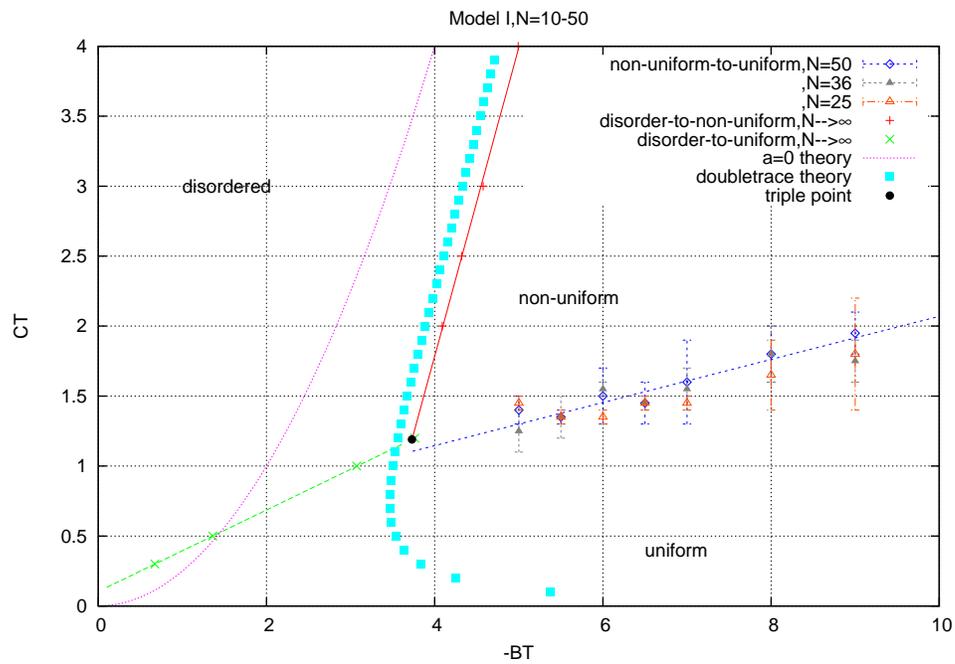}
\end{center}
\caption{The phase diagram of the multitrace matrix model of  \cite{O'Connor:2007ea}.  The Ising and matrix transition data points are not shown but we only indicate their extrapolated fits whereas the $N=25$, $N=36$ and $N=50$ stripe data points are included explicitly. 
}\label{pd}
\end{figure}

We consider as an example the multitrace matrix model of  \cite{O'Connor:2007ea} which comes with the $v$ values  $v_{2,1}=-1~,~v_{4,1}=3/2~,~v_{2,2}=0$. The action is given explicitly by
\begin{eqnarray}
V&=&{B}Tr M^2+{C}Tr M^4+D\big[ Tr M^2\big]^2+B^{'} (Tr M)^2+C^{'} Tr M Tr M^3.
\end{eqnarray}
The parameters $D$, $B^{'}$ and $C^{'}$ are constrained as $D=3N/4$, $B^{'}=\sqrt{N}/2$ and $C^{'}=-N$. The phase diagram of this model is computed by means of Monte Carlo elsewhere. The result is shown on figure (\ref{pd}). The details of the corresponding non-trivial lengthy Monte Carlo calculation will be reported elsewhere. As desired we have three stables phases in this particular model meeting at a triple point. In other words, we have established that this multitrace matrix model sustains the uniform ordered phase which is the first requirement. 

\subsection{Dimension from Critical Exponents}
The uniform ordered phase is also called the Ising phase precisely because we believe that the corresponding transition to the disordered phase is characterized by the universal critical exponents of the Ising model in two dimensions derived from the Onsager solution. These critical exponents are defined as usual by the following behavior 
   \begin{eqnarray}
&&m/N=<|Tr M|>/N \sim (B_c-B)^{\beta}\sim N^{-\beta/\nu}\nonumber\\
&&C_v/N^2 \sim (B-B_c)^{-\alpha}\sim N^{\alpha/\nu}\nonumber\\
&&\chi=<|Tr M|^2>-<|Tr M|>^2\sim (B-B_c)^{-\gamma}\sim N^{\gamma/\nu}\sim N^{2-\eta}\nonumber\\
&&\xi\sim|B-B_c|^{-\nu}\sim N.\label{cb}
\end{eqnarray} 
There are in total six critical exponents, the above five plus the critical exponent $\delta$ which controls the equation of state, but only two are truly independent because of the so-called scaling laws. The Onsager solution of the Ising model in two dimensions gives the following celebrated values \cite{Onsager:1943jn}

  \begin{eqnarray}
\nu=1~,~\beta=1/8~,~\gamma=7/4~,~\alpha=0~,~\eta=1/4~,~\delta=15.
\end{eqnarray} 
This fundamental result is very delicate to check explicitly in the Monte Carlo data. Since we must necessarily deal with the critical region we must face the two famous problems of finite size effects and critical slowing down. In this particular problem, the critical slowing down problem can be shown to start appearing in Monte Carlo simulations around $N>60$ so we will keep below this value and employ very large statistics of the order of $2^{20}$ to avoid it. A more systematic solution to this problem is to employ the Wolf algorithm \cite{Wolff:1988uh} which we do not attempt here. We simply employ here the ordinary Metropolis algorithm. The problem of finite size effects is also very serious for the measurement of the critical exponents since the above behavior (\ref{cb}) is supposed to hold only for large $N$. This problem can be avoided by not including values of $N$ less than $20$ and thus below we will quote for completeness $N=10$ and $N=15$ data but, in most cases, we will not take them into account in the fitting.

\begin{figure}[htbp]
\begin{center}
\includegraphics[width=5.0cm,angle=-90]{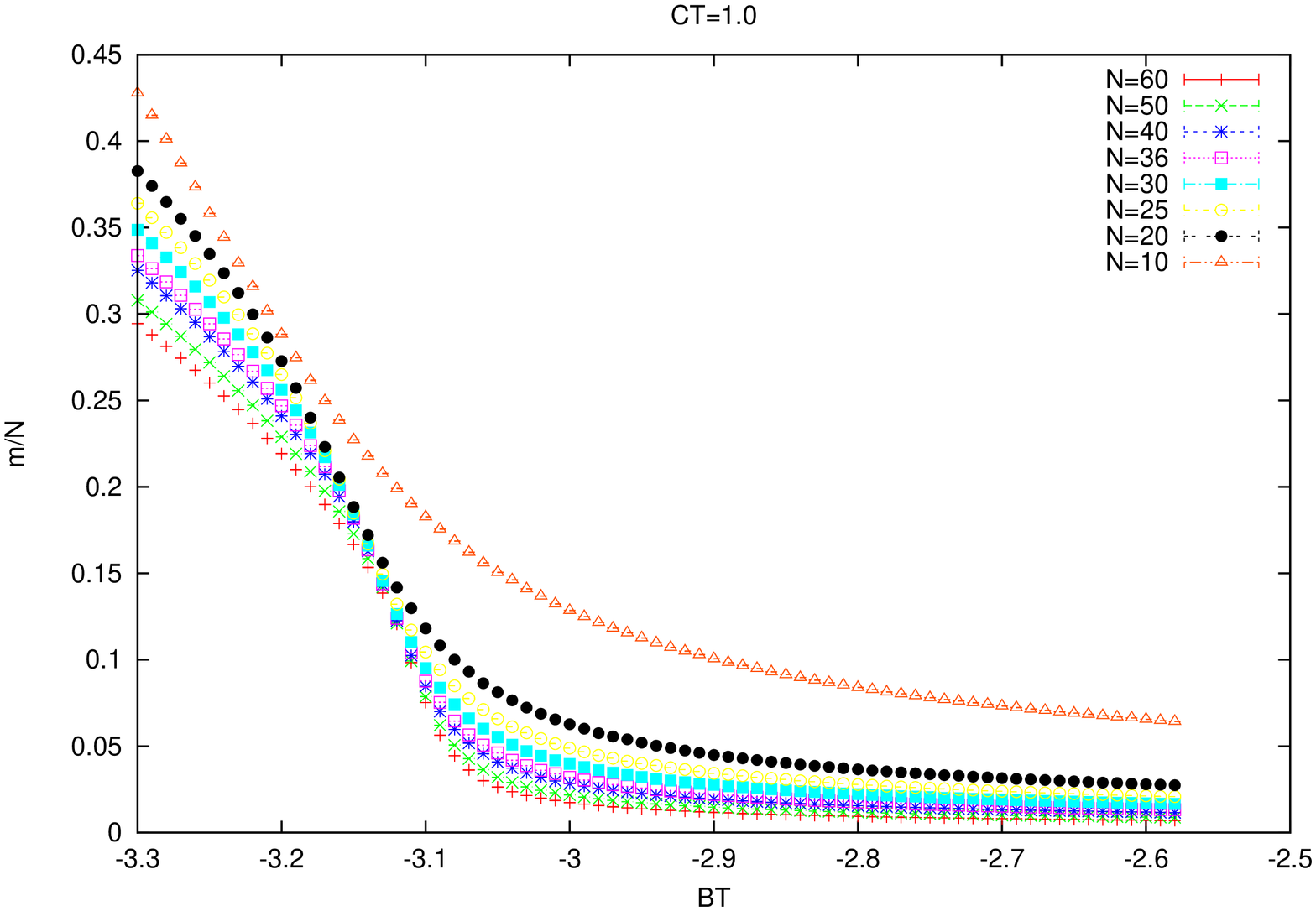}
\includegraphics[width=5.0cm,angle=-90]{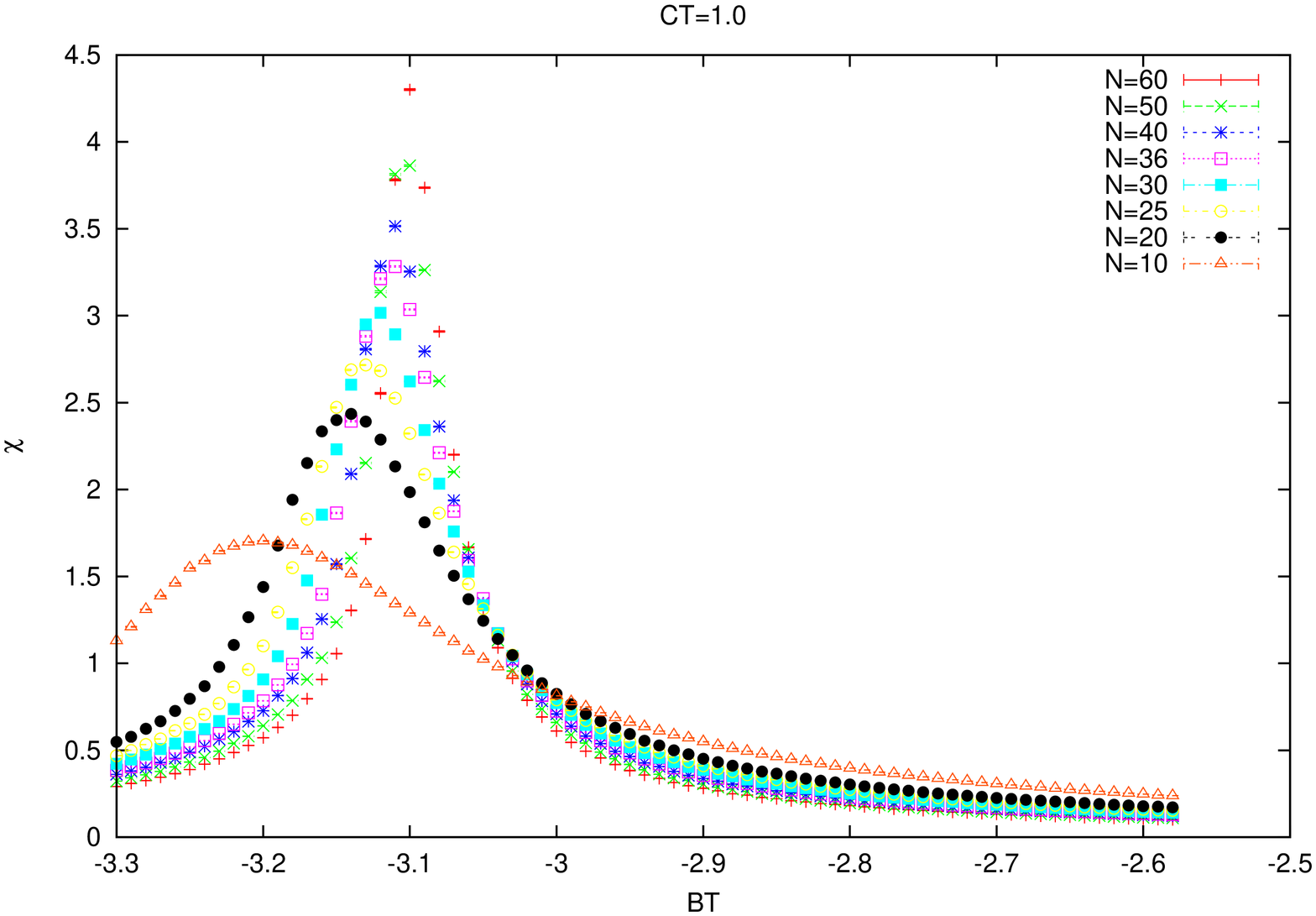}
\includegraphics[width=5.0cm,angle=-90]{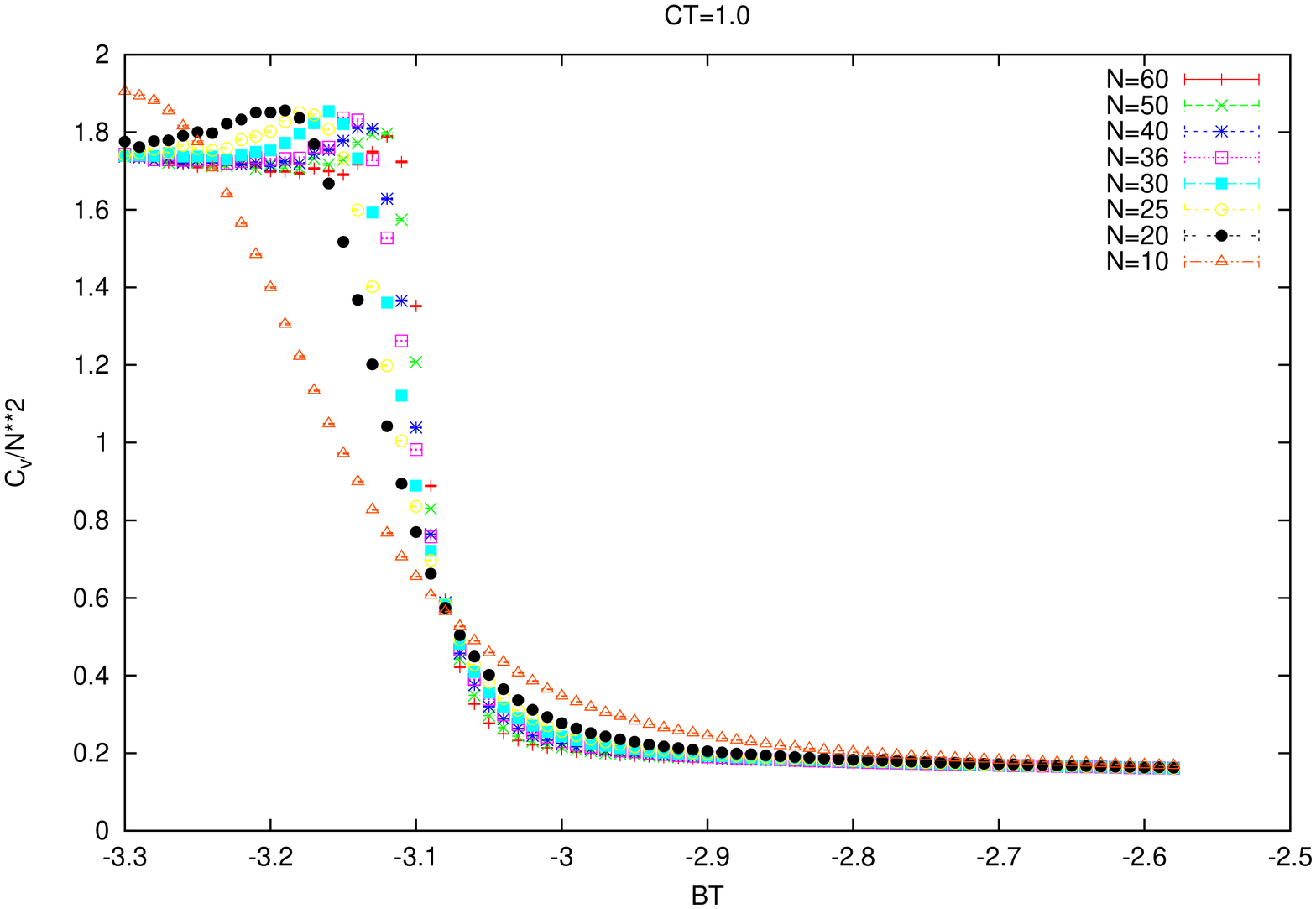}
\caption{The Ising critical behavior.}\label{critical}
\end{center}
\end{figure}

Since the Ising model appears  from the $\Phi^4$ theory for large values of the quartic coupling it is preferable to use values of $\tilde{C}$ as large as possible. However, we are limited from above by the appearance of the different physics of the transition between the disordered and non-uniform-ordered phases around $\tilde{C}=1.5$. Thus, we choose $\tilde{C}=1.0$ which is relatively large but well established to be within the Ising transition with an extrapolated critical point around $\tilde{B}= -3.07$ (see below). The critical behavior of the magnetization, susceptibility and specific heat around the critical value of  $\tilde{B}= -3.10$ is shown on figure (\ref{critical}). We attach in table (\ref{tablehis4}) some data relevant for the computation of the critical exponents $\nu$, $\beta$, $\gamma$ and $\alpha$. The other critical exponents can be determined via scaling laws. 

\begin{figure}[htbp]
\begin{center}
\includegraphics[width=5.0cm,angle=-90]{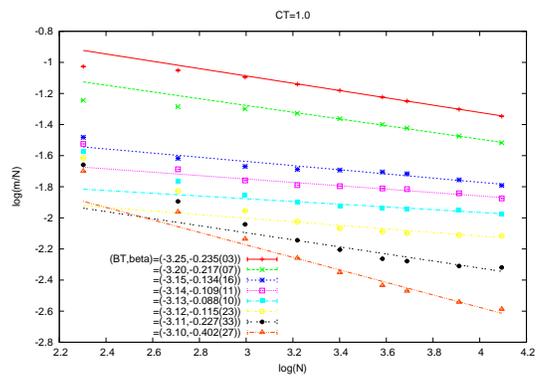}
\caption{The critical exponent $\beta$.}\label{critical1}
\end{center}
\end{figure}

\begin{figure}[htbp]
\begin{center}
\includegraphics[width=5.0cm,angle=-90]{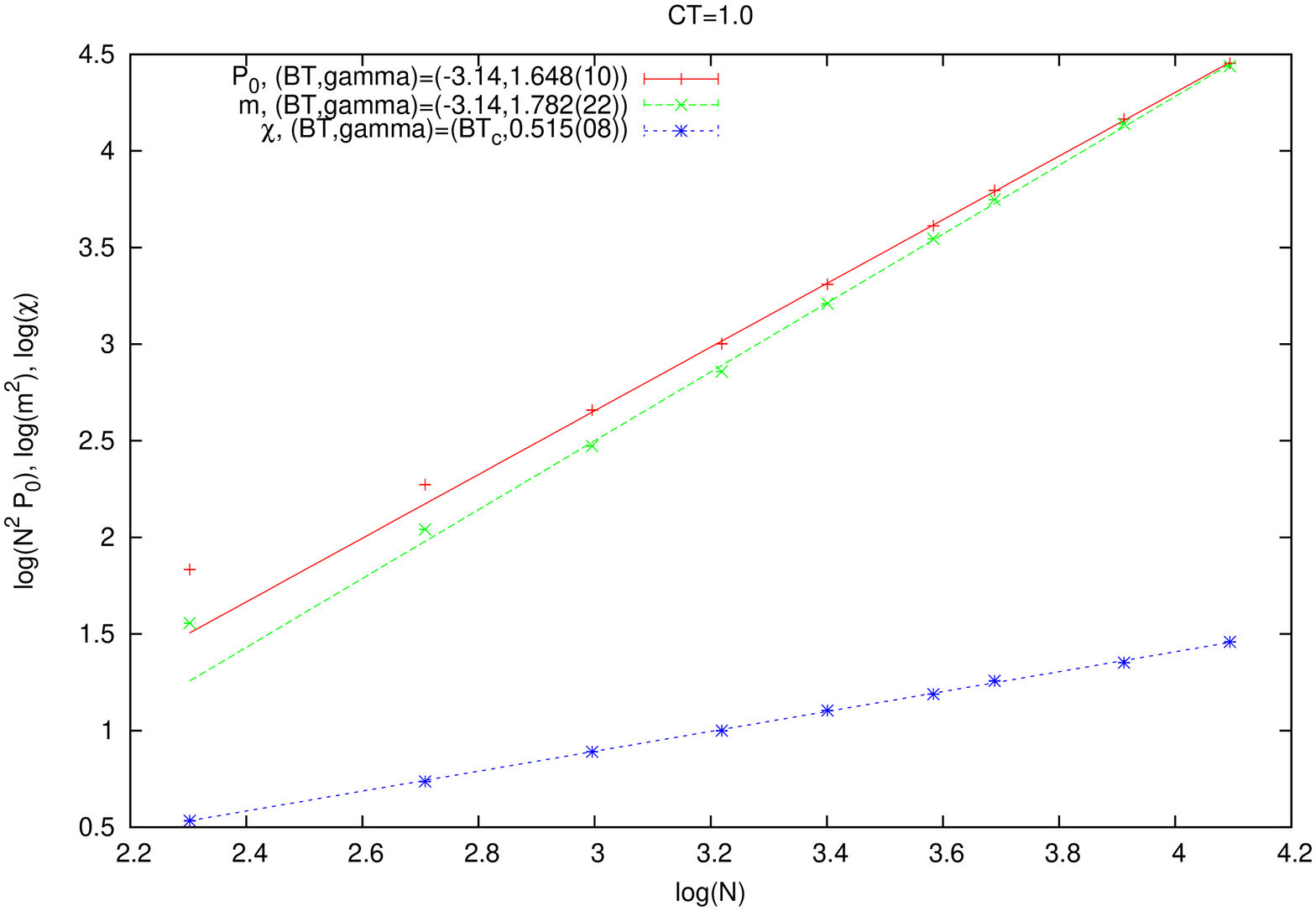}
\includegraphics[width=5.0cm,angle=-90]{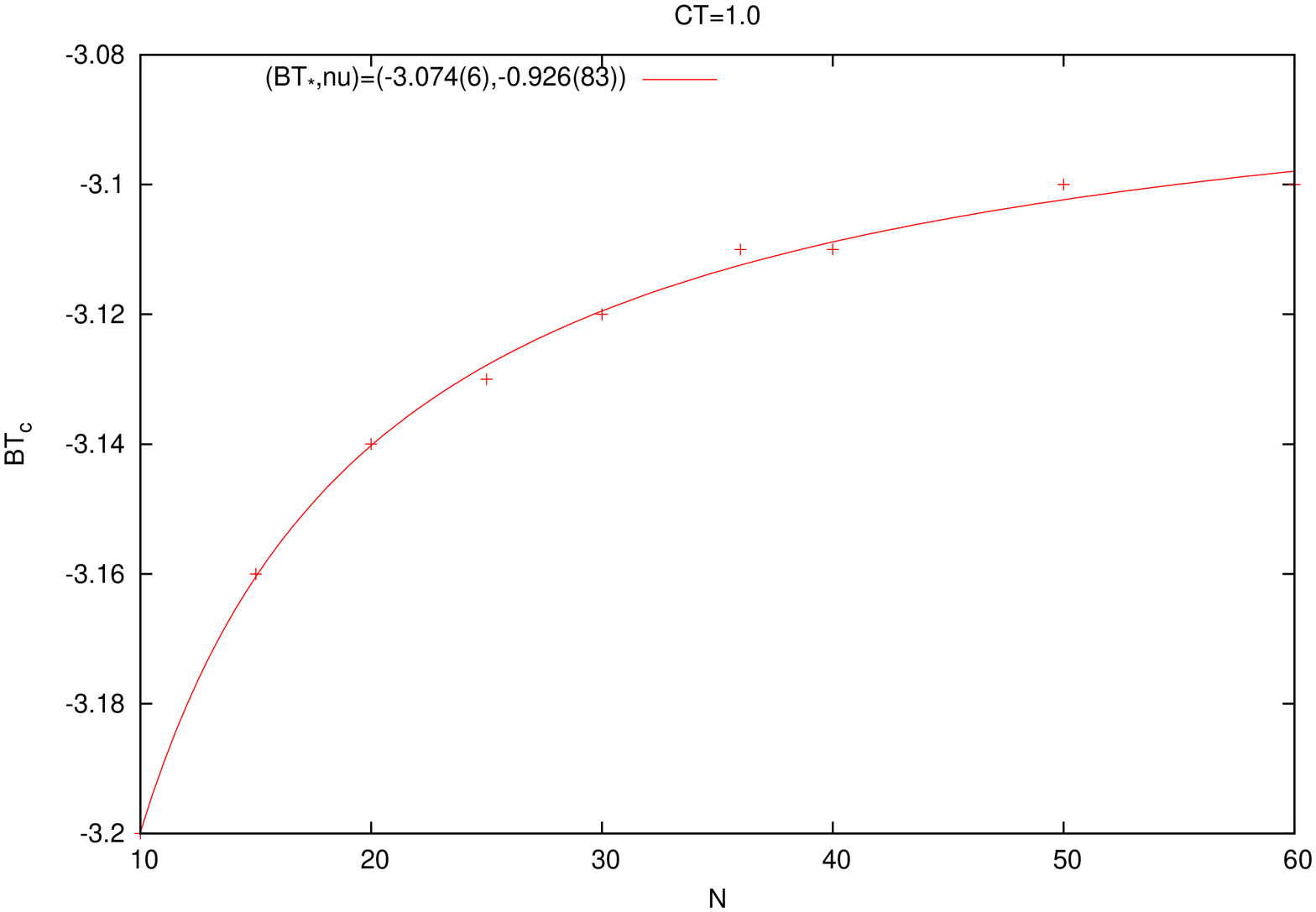}
\includegraphics[width=5.0cm,angle=-90]{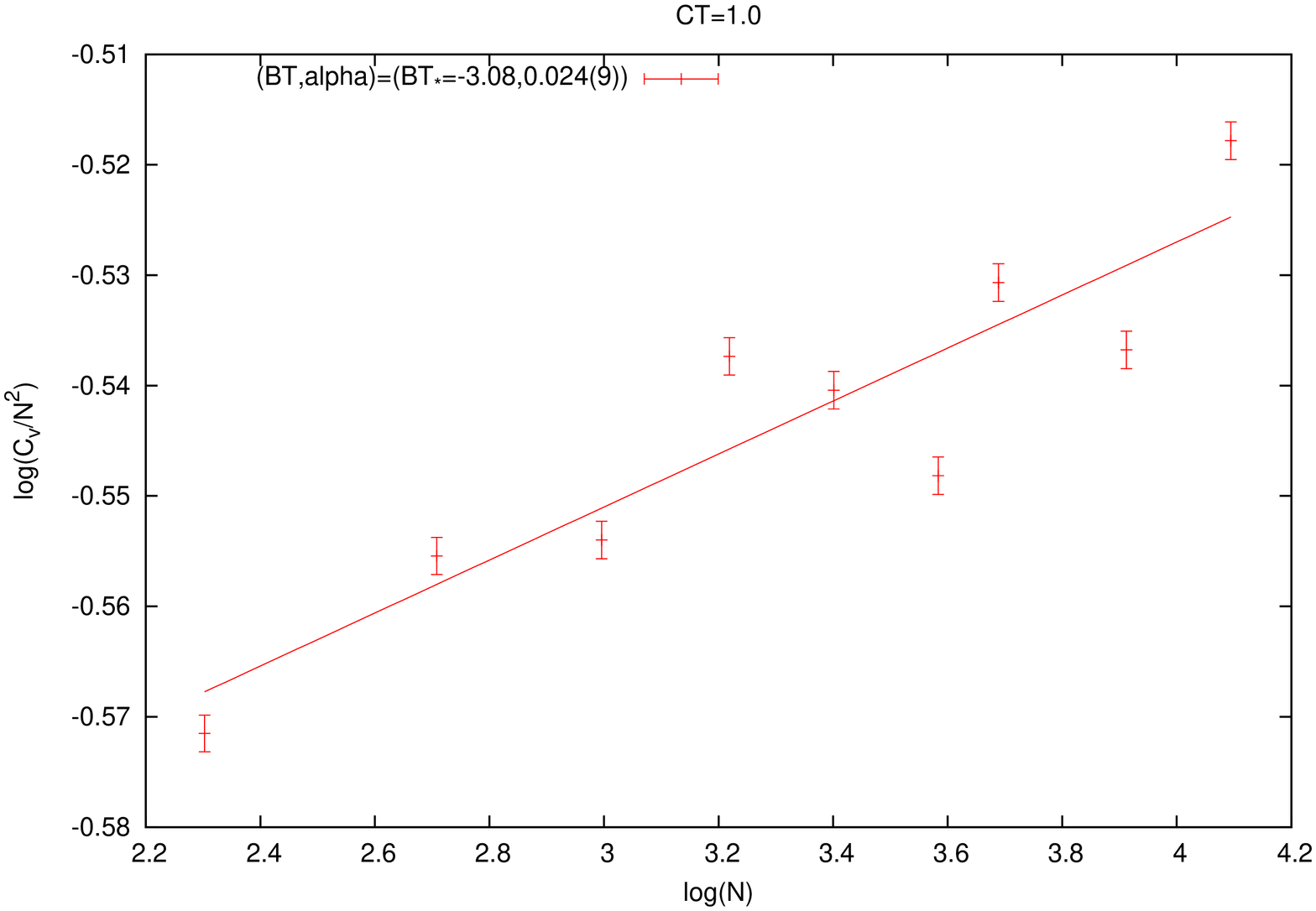}
\caption{The critical exponents $\nu$, $\gamma$ and $\alpha$.}\label{critical2}
\end{center}
\end{figure}

The measurements of the critical exponents $\nu$, $\beta$, $\gamma$ and $\alpha$ proceeds as follows:
\begin{itemize}
\item{\bf Critical Point and The Critical Exponent $\nu$:} By plotting the critical point $\tilde{B}_c$ obtained for each $N$ versus $N$ (first and second columns of  table (\ref{tablehis4})) we get immediately both the $N=\infty$ critical point and the critical exponent $\nu$. We obtain (see figure (\ref{critical2}))
  \begin{eqnarray}
&&\tilde{B}_c=-1.061(168).N^{-0.926(83)}-3.074(6)\Rightarrow~,~\nu=0.926(83).\label{nu}
\end{eqnarray} 
Also we obtain
 \begin{eqnarray}
&& \tilde{B}_*=-3.074(6).\label{me1}
\end{eqnarray} 
This prediction for $\nu$ agrees reasonably well with the Onsager calculation. In the following we will assume for simplicity that $\nu=1$. The above fit is the only instance in which we have included $N=10$ and $N=15$ and thus we believe that the obtained value of $-\tilde{B}_*$ is an underestimation of the true critical point. 
\item {\bf Magnetization and The Critical Exponent $\beta$:} The magnetization and the zero power are defined by
\begin{eqnarray}
m=<|Tr M|>~,~\chi=<|Tr M|^2>-<|Tr M|>^2.
\end{eqnarray}
\begin{eqnarray}
P_0=<\big(\frac{1}{N}Tr M)^2>.
\end{eqnarray}
Measurements of the magnetization $m/N$ were performed near the extrapolated critical point $\tilde{B}=-3.07$ for $\tilde{C}=1.0$ but inside the uniform ordered phase. These are then used to compute the critical exponent $\beta$ by searching for a power law behavior. 

More precisely, we measure $\ln(m/N)$ versus $\ln N$ for each value of $\tilde{B}$ very near and around  $\tilde{B}=-3.10$, fit to a straight line in the range $20\leq N\leq 60$ and compute the slope $\beta$, then search for the flattest line, i.e. the smallest slope $\beta$. This value marks the transition from the Ising phase to the disordered phase. Deeep inside the Ising phase the slope should approach the mean field value $-1/4$ which can be shown from the scaling behavior of the dominant configuration. After determining the critical value we then consider the value of $\tilde{B}$ nearest to it but within the Ising phase and take the slope there to be the value of the critical exponent $\beta$. In our example here, the flattest line occurs at $\tilde{B}=-3.13$ with slope $-0.088(10)$ after which the slope becomes $-0.109(11)$ at $\tilde{B}=-3.14$. The slope goes fast to the mean field value $-0.25$ as we keep decreasing $\tilde{B}$. See figure (\ref{critical1}). Our measured value of the critical point $\tilde{B}_*$ from the magnetization and of the critical exponent $\beta$ are therefore
 \begin{eqnarray}
&& \tilde{B}_*=-3.13.\label{me2}
\end{eqnarray}  
  \begin{eqnarray}
\ln\frac{m}{N}=-0.109(11).\ln N-1.423(43)\Rightarrow \beta=-0.109(11).\label{beta}
\end{eqnarray} 
\item {\bf Susceptibility and Zero Power and The Critical Exponent $\gamma$:}  The measurement of the critical exponent $\gamma$ is quite delicate and will be done indirectly as follows. We rewrite the susceptibility in terms of the zero power and magnetization as
 \begin{eqnarray}
\chi&=&<|Tr M|^2>-<|TrM|>^2\nonumber\\
&=&N^2P_0-m^2.
\end{eqnarray} 
The critical exponent $\gamma$ in terms of the critical exponent $\gamma^{'}$ of $P_0$ is then given by
 \begin{eqnarray}
\gamma=2+\gamma^{'}.
\end{eqnarray} 
By using the results shown on table (\ref{tablehis4}) at $\tilde{B}=-3.14$, plotted on figure (\ref{critical2}), we obtain the following exponents
   \begin{eqnarray}
\ln P_0=-0.352(10).\ln N-2.289(36)\Rightarrow \gamma^{'}=-0.352(10).
\end{eqnarray} 
Or equivalently 
\begin{eqnarray}
\ln N^2 P_0=1.648(10).\ln N-2.289(36)\Rightarrow \gamma=1.648(10).
\end{eqnarray} 
For consistency we can check that the second term in the  susceptibility behaves using the result (\ref{beta}) as
 \begin{eqnarray}
\ln m^2=1.782(22).\ln N-2.846(86)\Rightarrow \gamma=1.782(22).
\end{eqnarray} 
Our two measurements of the critical exponent $\gamma$ agree reasonably well with the Onsager values. 

If we try to fit the values of the susceptibility at its maximum shown in third column of table  (\ref{tablehis4}), i.e. at the peak which keeps slowly moving with $\tilde{B}$, then we will obtain a very bad underestimate of the critical exponent $\gamma$ given by
   \begin{eqnarray}
\ln \chi_{\rm max}=0.515(08).\ln N-0.652(30)\Rightarrow \gamma=0.515(08).
\end{eqnarray}  
This in our mind is due in part to the dependence of $\tilde{B}_c$ on $N$ and in another part is an indication of the critical slowing down problem showing up in the measurement of this second moment, i.e. the size of the fluctuations is observed to grow with $N$ at the critical point but not at the correct rate indicated by the independent measurements of the zero moment and the magnetization. See figure (\ref{critical2}).
\item {\bf Specific Heat and The Critical Exponent $\alpha$:} The sepcific heat is defined by
\begin{eqnarray}
C_v=<S^2>-<S>^2.
\end{eqnarray}
The critical point $\tilde{B}_*$ as measured from the specific heat is identified by the intersection point of the various curves with different $N$ shown on figure (\ref{critical}). We get
 \begin{eqnarray}
&& \tilde{B}_*=-3.08.
\end{eqnarray}  
This measurement is contrasted very favorably with the independent measurement obtained from the extrapolated value of $\tilde{B}_c$ shown in equation (\ref{me1}) but should also be contrasted with the measurement obtained from the magnetization shown in equation (\ref{me2}).

By using the results shown on table (\ref{tablehis4}) at the critical point $\tilde{B}=-3.08$, plotted on figure (\ref{critical2}), we obtain the following exponent
   \begin{eqnarray}
\ln \frac{C_v}{N^2}=0.024(9).\ln N-0.623(31)\Rightarrow \alpha=0.024(9).
\end{eqnarray} 
\end{itemize}
\begin{table}[H]
\begin{center}
\begin{tabular}{ |c|c|c|c|c|c|c| }  
 \hline
 $N$ & $\tilde{B}_c,\tilde{B}_*=-3.07$ &  $ \chi_c$ &  $ (C_v)_*,\tilde{B}_*=-3.08$ & $\tilde{B}< \tilde{B}_*=-3.13$  &$m_{<*}$ &  $10^3(P_0)_{<*} $ \\ 
 \hline
 $10$ & $- 3.20  $ & $1.704(2)$  & $56.467(94) $& $- 3.14  $ & $  2.1776(12) $&  $6.256(6)$   \\
\hline
 $15$ &$- 3.16  $ & $2.089(2)$  & $   129.111(217)        $& $- 3.14  $& $  2.7750(14)  $  &  $4.315(4)$  \\
\hline
 $20$ &$- 3.14  $ & $2.436(3)$& $ 229.861(389)   $& $- 3.14  $&$  3.4423(15)   $&  $3.571(2)$ \\
 \hline
$25$ & $-3.13 $  &$2.716(3) $ & $  365.183(621)  $& $- 3.14  $ &$  4.1759(16) $&   $ 3.220(2)$ \\
 \hline
$30$ &$-3.12 $  &  $ 3.017(4) $ & $ 524.253(891)$& $- 3.14  $ &$ 4.9772(16) $&  $3.042(2)$ \\
 \hline
$36$ &$-3.11 $  & $3.283(4) $ & $ 749.099(1267)$& $- 3.14  $ &$ 5.8878(15)  $&  $2.860(1)$  \\
 \hline
$40$ &  $-3.11 $ & $3.515(4) $ & $  941.139(1607) $& $- 3.14  $&$6.5134(14)   $ & $2.782(1)$  \\
 \hline
$50$ &  $-3.10 $ &$3.864(4) $  & $  1461.597(2479) $& $- 3.14  $ &$ 7.9250(12)   $& $2.576(1)$   \\
 \hline
$60$ &  $-3.10 $ & $4.301(5)$& $ 2144.929(3658) $& $- 3.14  $&$9.2021(11)  $ &  $2.388(1)$ \\
\hline
\end{tabular}
\end{center}
\caption{Measurements of the magnetization $(m/N)_{<*}$, the susceptibility $\chi_{<*}$, via the zero power $(P_0)_{<*}$, and the specific heat $(C_v/N^2)_{*}$  used to compute the critical exponents $\beta$, $\gamma$ and $\alpha$ respectively. Here  $\tilde{C}=1.0$, the extrapolated critical point is $\tilde{B}=-3.07$, the critical point as intersection point of curves of specific heat is $\tilde{B}=-3.08$, and the critical point as the flattest line of decrease of magnetization is $\tilde{B}=-3.13$.}\label{tablehis4}
\end{table}
\subsection{Free Propagator from Wigner Semicircle Law}
We can also measure the emergent geometry by measuring the free propagator of the theory. This will give us information on both the dimension and the metric  since the free propagator is the inverse of the Laplacian  $\Delta$ which fully encodes the underlying geometry in the sense of  \cite{Connes:1994yd,Frohlich:1993es}. This goes as follows \cite{Steinacker:2005wj}.

\begin{figure}[htbp]
\begin{center}
\includegraphics[width=5.0cm,angle=-90]{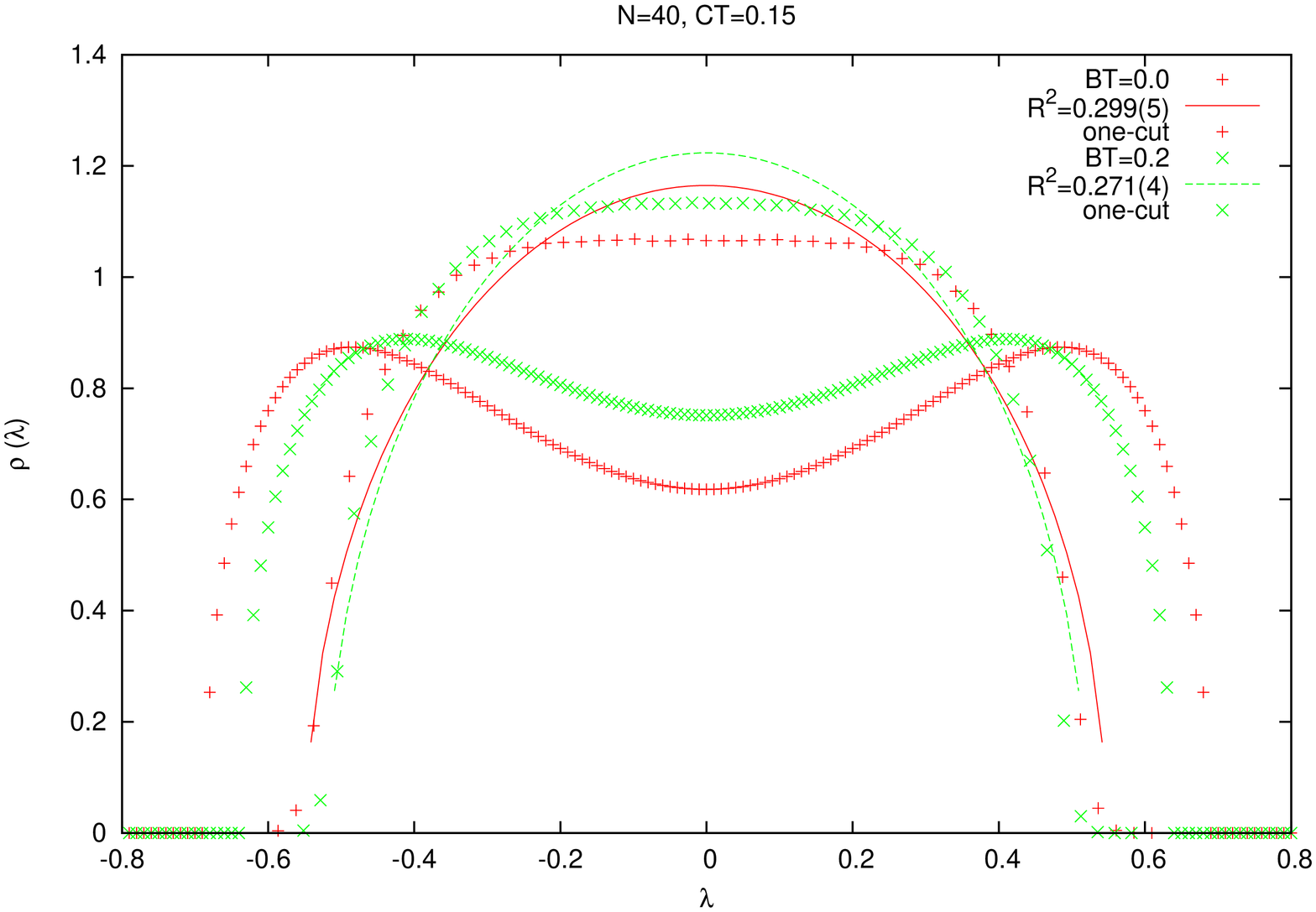}
\includegraphics[width=5.0cm,angle=-90]{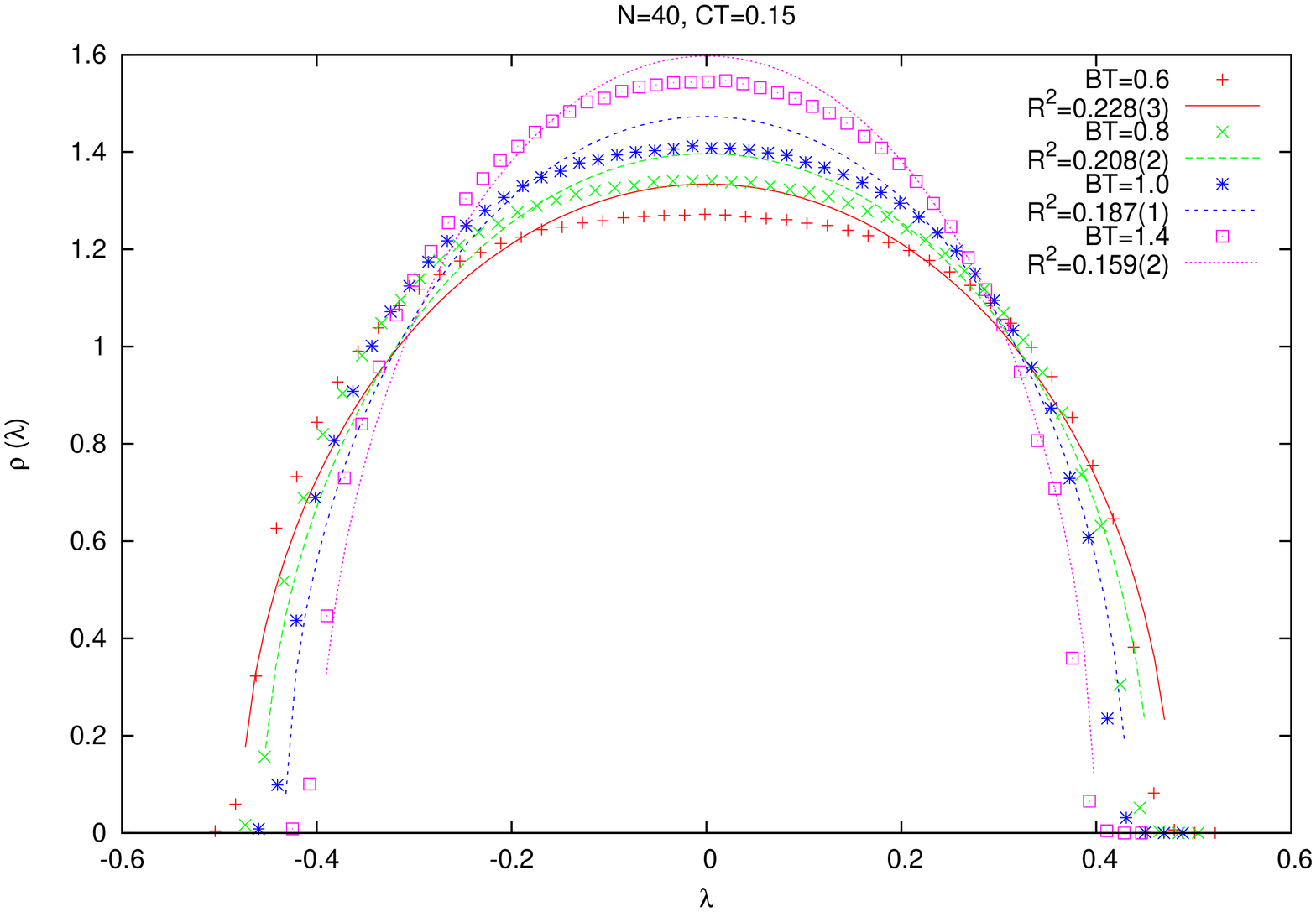}
\includegraphics[width=5.0cm,angle=-90]{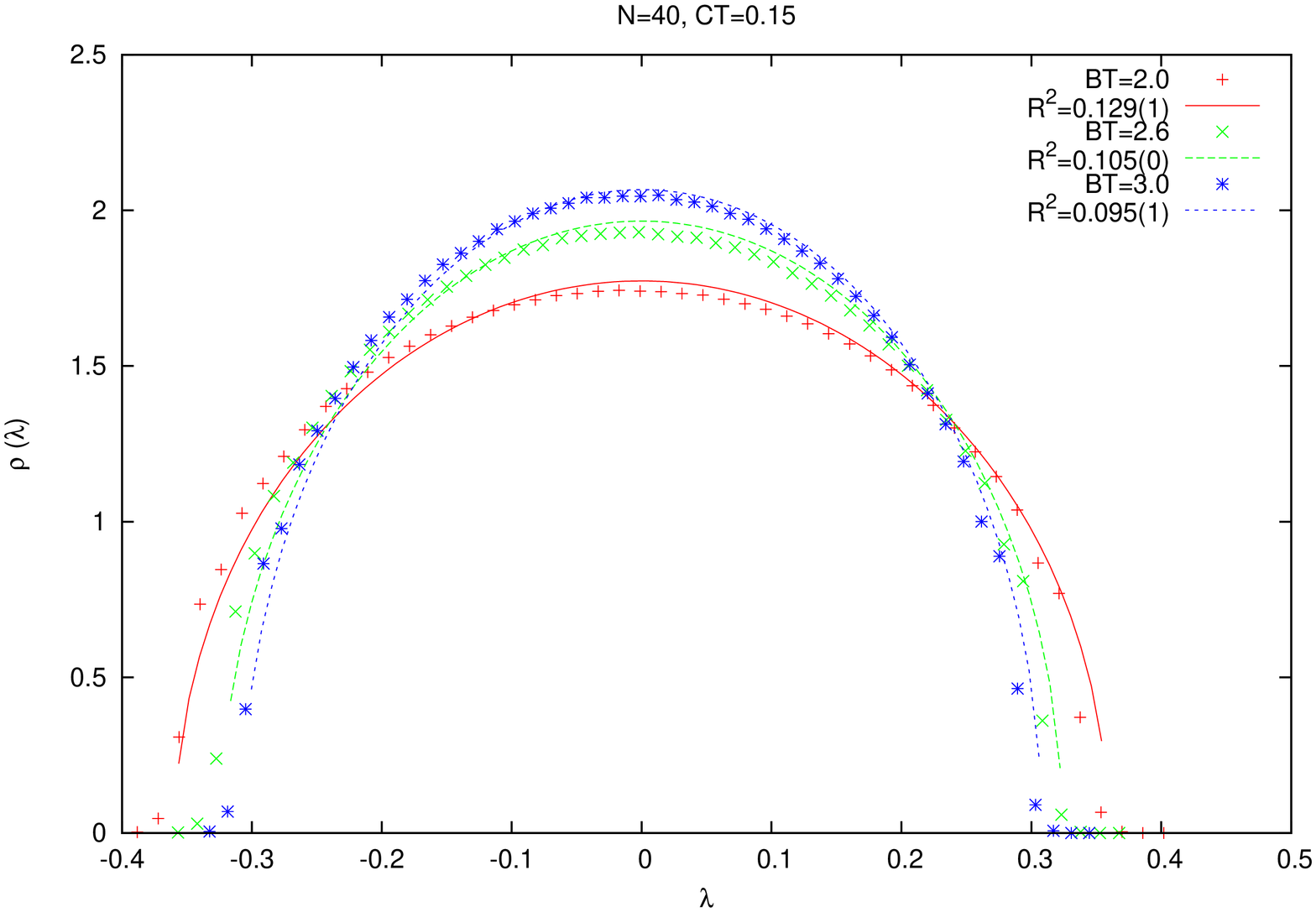}
\includegraphics[width=5.0cm,angle=-90]{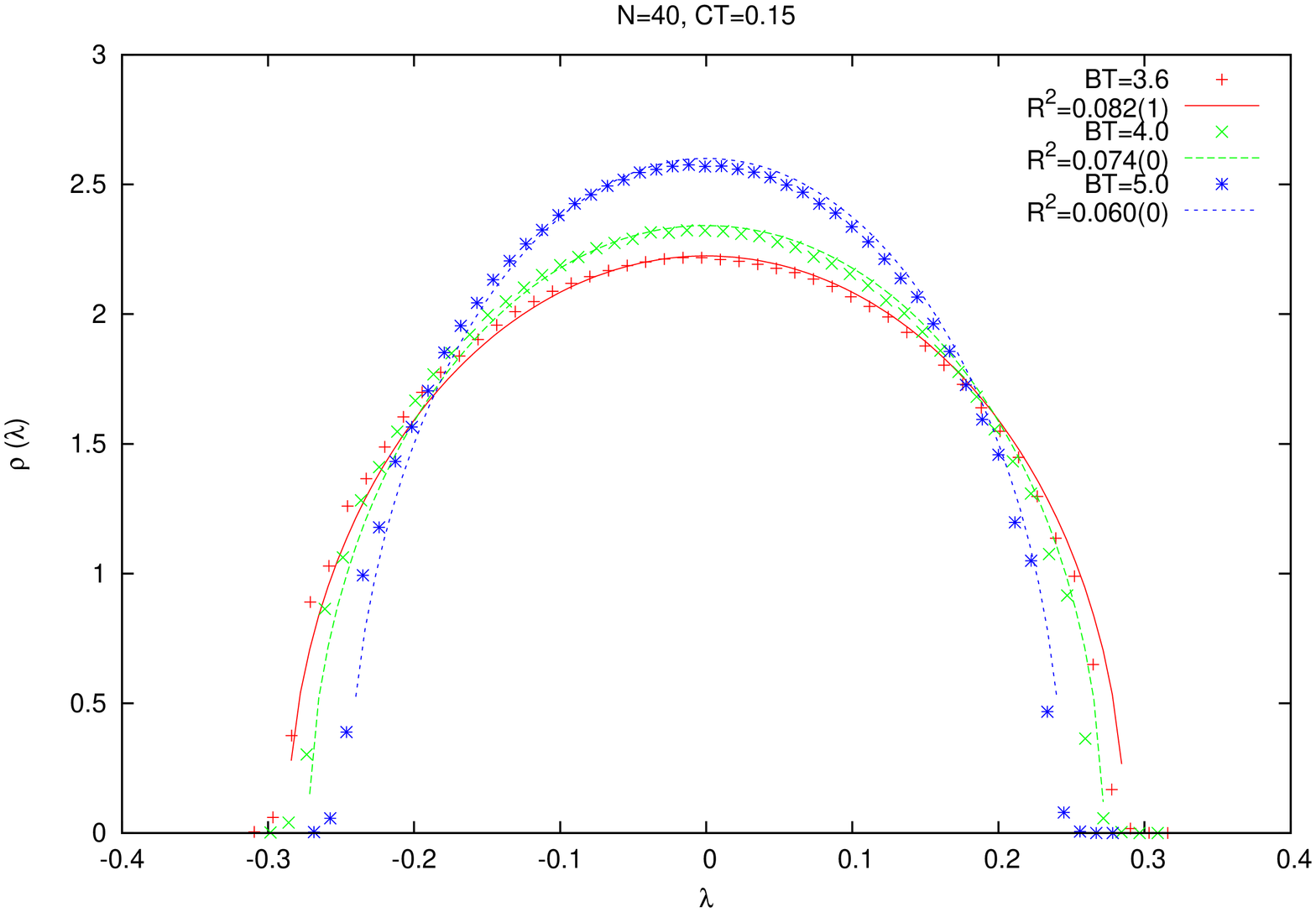}
\caption{The semicircle  law as a function of $\tilde{B}$.}\label{Wigner0}
\end{center}
\end{figure}

\begin{figure}[htbp]
\begin{center}
\includegraphics[width=5.0cm,angle=-90]{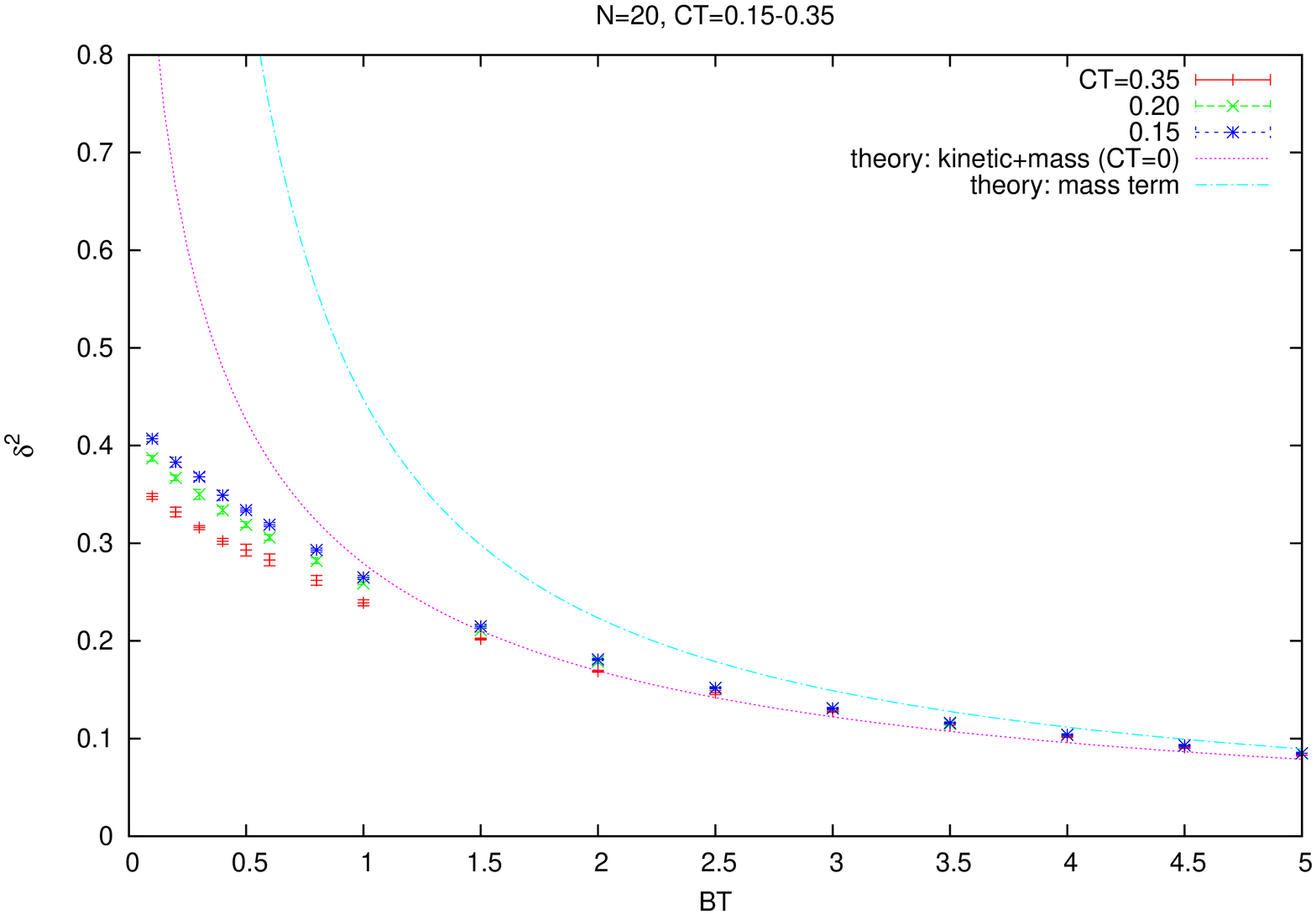}
\includegraphics[width=5.0cm,angle=-90]{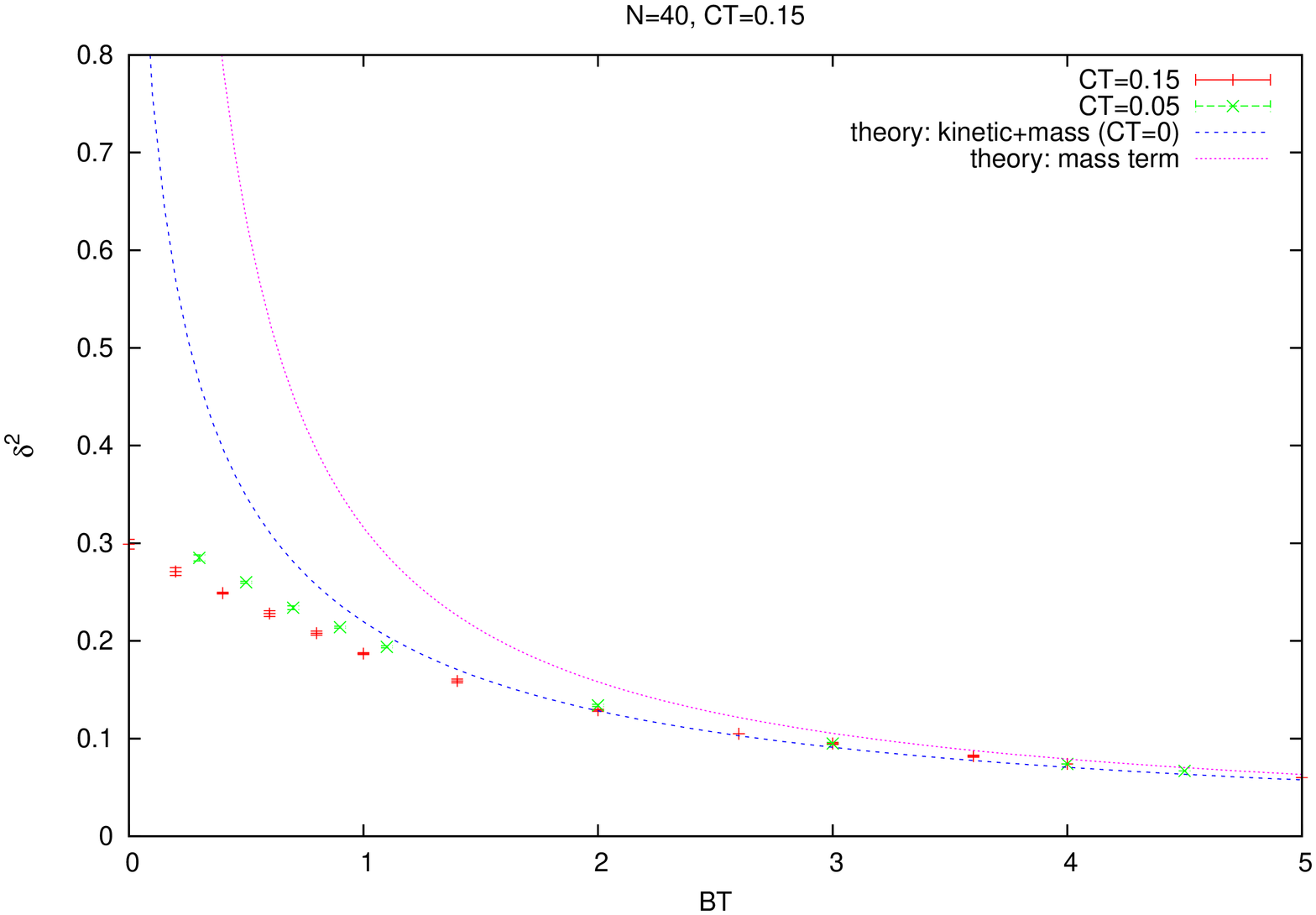}
\caption{The behavior of the  radius of the Wigner semicircle law $\delta^2=\alpha_0^2$ as a function of $\tilde{B}$.}\label{Wigner}
\end{center}
\end{figure}

A noncommutative phi-four on a $d-$dimensional noncommutative Euclidean spacetime ${\bf R}^d_{\theta}$ reads in position representation
\begin{eqnarray}
S&=&\int d^dx \big(\frac{1}{2}\partial_i \Phi\partial_i {\Phi}+\frac{1}{2}m^2{\Phi}^2+\frac{\lambda}{4}{\Phi}_*^4\big).
\end{eqnarray} 
The first step is to regularize this theory in terms of  a finite ${\cal N}-$dimensional matrix $\Phi$ and rewrite the theory in matrix representation. Then we diagonalize the matrix $\Phi$. The measure becomes $\int \prod_i d\Phi_i\Delta^2(\Phi)\int dU$ where $\Phi_i$ are the eigenvalues, $\Delta^2(\Phi)=\prod_{i<j}(\Phi_i-\Phi_j)^2$ is the Vandermonde determinant and $dU$ is the Haar measure. The effective probability distribution of the eigenvalues $\Phi_i$ can be determined uniquely from the behavior of the expectation values $<\int  d^dx \Phi_*^{2n}(x)>$. These objects clearly depend only on the eigenvalues $\Phi_i$ and are computed using a sharp UV cutoff $\Lambda$. If we are only interested in the eigenvalues of the scalar matrix $\Phi$ then the free theory $\lambda=0$ can be replaced by the effective matrix model \cite{Steinacker:2005wj}
\begin{eqnarray}
S=\frac{2{\cal N}}{\alpha_0^2}Tr\Phi^2.
\end{eqnarray} 
This result can be traced to the fact that planar diagrams dominates over the non-planar ones in the limit $\Lambda\longrightarrow\infty$. This means in particular that the eigenvalues $\Phi_i$ are distributed according to the famous Wigner semi-circle law with $\alpha_0$ being the largest eigenvalue, viz
\begin{eqnarray}
\rho(t)=\frac{2}{\pi\alpha_0^2}\sqrt{\alpha_0^2-t^2}~,~-\alpha_0\leq t\leq +\alpha_0.\label{wi}
\end{eqnarray} 
In the most important cases of $d=2$ and $d=4$ dimensions we have explicitly 
\begin{eqnarray}
\alpha_0^2(m,\Lambda)=\frac{1}{4\pi^2}\big(\Lambda^2-m^2\ln(1+\frac{\Lambda^2}{m^2})\big)~,~d=4.
\end{eqnarray} 
\begin{eqnarray}
\alpha_0^2(m,\Lambda)=\frac{1}{\pi}\ln(1+\frac{\Lambda^2}{m^2})~,~d=2.\label{92}
\end{eqnarray} 
Obviously, dimension four is eliminated by the results of the critical exponents. In two dimensions the regulator $\Lambda$ originates in only one of two possible noncommutative spaces  \cite{Steinacker:2005wj}: 
\begin{enumerate}
\item {\bf Fuzzy Torus:} As it turns the results on the fuzzy torus are different from those obtained using a sharp momentum cutoff due to the different behavior of the propagator for large momenta and as a consequence the resulting formula for  $\alpha_0^2$ is different from the above equation (\ref{92}). We obtain instead
\begin{eqnarray}
\alpha_0^2(m,\Lambda)=4\int_0^{\pi}\frac{d^2r}{(2\pi)^2}\frac{1}{\sum_i(1-\cos r_i)+m^2l^2/2}~,~d=2.
\end{eqnarray}
$l$ here is the lattice spacing, the noncommutativity is quantized as $\theta={Nl^2}/{\pi}$ and the cutoff is
\begin{eqnarray}
\Lambda=\frac{\pi }{l}=\sqrt{\frac{N\pi}{\theta}}.
\end{eqnarray}
The above behavior can be easily excluded in our Monte Carlo data and by hindsight we know that this should be indeed so because the original multitrace approximation is relevant to the fuzzy sphere.
\item {\bf Fuzzy Sphere:} The fuzzy sphere ${\bf S}_N^2={\bf CP}_N^1$ is the simplest of fuzzy projective spaces ${\bf CP}_N^n$. In this case ${\cal N}=N+1$ and the scalar field $\Phi$ becomes an $N\times N$ matrix $\phi$ given by $\phi=\sqrt{2\pi/Na}\Phi$. In this case the cutoff is given in terms of the matrix size $N$ and the radius $R$ of the sphere by
\begin{eqnarray}
\Lambda=\frac{N}{R}.
\end{eqnarray}
Also, in this case the mass parameters $B$ and $m^2$ are related by
 \begin{eqnarray}
m^2=\frac{b}{aR^2}.
\end{eqnarray}
By using $\tilde{B}=B/N^{3/2}$ and choosing $a=2\pi/N$, so that $\Phi=\phi$, we obtain
 \begin{eqnarray}
\frac{\Lambda^2}{m^2}=\frac{2\pi}{\sqrt{N}\tilde{B}}.
\end{eqnarray}
We get then
\begin{eqnarray}
\alpha_0^2(m,\Lambda)=\frac{1}{\pi}\ln(1+\frac{2\pi}{\sqrt{N}\tilde{B}}).\label{theo}
\end{eqnarray} 
In the limit $B\longrightarrow \infty$ we get the one-cut $\delta^2=2N/B$ of the Gaussian matrix model $B Tr M^2$, viz $B=2{\cal N}/{\alpha}_0^2$. This can also be obtained by taking the limit  $B\longrightarrow \infty$ of the one-cut (deformed Wigner semicircle law) solution 
\begin{eqnarray}
\rho(\lambda)&=&\frac{1}{N\pi}(2C\lambda^2+B+C\delta^2)\sqrt{\delta^2-\lambda^2}~,~\delta^2=\frac{1}{3C}(-B+\sqrt{B^2+12 NC})\label{1cut}\nonumber\\
\end{eqnarray}
of the quadratic matrix model  $B Tr M^2+C Tr M^4$.

This result was also generalized in \cite{Nair:2011ux}. The eigenvalues distribution of a free scalar field theory on the fuzzy sphere with an arbitrary kinetic term, viz $S=Tr(M{\cal K}M+BM^2)/2$, where ${\cal K}(0)=0$ and ${\cal K}$ is diagonal in the basis of polarization tensors $T_m^l$,  is always given by a Wigner semicircle law with a radius
\begin{eqnarray}
R^2=\delta^2&=&\alpha_0^2=\frac{4f(B)}{N}~,~f(B)=\sum_{l=0}^{N-1}\frac{2l+1}{{\cal K}(l)+B}.
\end{eqnarray}
\end{enumerate}
Some Monte Carlo results are shown on figures (\ref{Wigner0}) and (\ref{Wigner}). These are obtained in Monte Carlo runs with $2^{20}$ thermalization steps and $2^{18}$ thermalized configurations where each two configurations are separated by $2^4$ Monte Carlo steps in order to reduce auto-correlation effects. We consider $N=20-40$, $\tilde{C}=0.05-0.35$ and $\tilde{B}=0-5$. 

It is not difficult to convince ourselves that the mass parameter $B$ is precisely the mass squared  in this regime. 
 For each value of $(N,\tilde{C},\tilde{B})$ we compute the eigenvalues distribution $\rho(\lambda)$ and fit it to the Wigner semicircle law  (\ref{wi}) (see figure  (\ref{Wigner0})). We obtain thus a measurement of the radius of the Wigner semicircle law $\delta^2=\alpha_0^2=R^2$. We have checked carefully that in this regime the Wigner semicircle law is the appropriate behavior rather than the one-cut solution (\ref{1cut}) as evidenced by the first graph in figure  (\ref{Wigner0}). The measurement of the radii  $\delta^2$ for various values of $\tilde{B}$ is then plotted and compared with the expected theoretical behaviors (\ref{theo}) as well as with the $B\longrightarrow \infty$ behavior $\delta^2=2N/B$  (see figure  (\ref{Wigner})). The agreement with  (\ref{theo}) is very reasonable with some deviation for small values of $\tilde{B}$ as we approach the non-perturbative region where the uniform ordered phase appears at some $\tilde{B}<0$. This discrepancy for small values of $\tilde{B}$ is already seen on figure  (\ref{Wigner0}) when we fit the distributions to the Wigner semicircle law. However, this effect is reduced as we decrease the value of $\tilde{C}$.

In summary we conclude that we are indeed dealing with the geometry of the fuzzy sphere and, given hindsight, we know that this should be true.

\section{Generalization and Conclusion}
The emergence of geometry in the very early universe is a problem of fundamental importance to our understanding of quantum gravity and cosmology. In this letter, we have proposed a novel scenario for the emergence of geometry in random multitrace matrix models which depend on a single hermitian matrix $M$ with full unitary $U(N)$ invariance and without any kinetic term. Thus, the model under consideration has no geometry a priori precisely because of the absence of a kinetic term. On the other hand, previous proposals of emergent geometry required the input of several matrices with some rotational symmetry group besides the $U(N)$ gauge symmetry \cite{DelgadilloBlando:2007vx}. 

Our proposal consists in checking whether or not the uniform ordered phase is sustained by the multitrace matrix model under consideration. If yes, then the dimension of the underlying geometry, in the region of the phase diagram where the uniform ordered phase is stable, can be inferred from the values of the critical exponents of the Ising phase transition.  Whereas, the metric/Laplacian of this geometry can be inferred from the behavior of the free propagator encoded in the Wigner semicircle law behavior of the eigenvalues distribution of the matrix $M$ in the weakly coupled regime. An explicit example is given in which the geometry of the fuzzy sphere emerges, with all the correct properties, in the phase diagram of a particular multitrace matrix model containing multitrace terms depending on the moments $m_1=Tr M$, $m_2=TrM^2$ and $m_3=Tr M^3$ in a particular way  \cite{O'Connor:2007ea}.

This idea can be generalized in a straightforward way to all higher fuzzy projective spaces ${\bf CP}^n$ and fuzzy tori ${\bf T}^n$ by tuning appropriately the coefficients of the multitrace matrix model and/or including higher moments in the multitrace matrix model .


\paragraph{Acknowledgment:}  
This research was supported by CNEPRU: "The National (Algerian) Commission for the Evaluation of
University Research Projects"  under contract number ${\rm DO} 11 20 13 00 09$.

\end{document}